\documentclass[12pt]{article}
\usepackage{epsfig,cite}
\newcommand{\p}{\mbox{e$^+$}}
\newcommand{\n}{\mbox{e$^-$}}
\newcommand{\pn}{\mbox{e$^+$e$^-$}}
\newcommand{\UTa}{\mbox{$^{238}$U + $^{181}$Ta\ }}

\newcommand{\UPb}{\mbox{$^{238}$U + $^{206}$Pb\ }}

\begin{document}
\begin{titlepage}
\vspace*{3mm}
\begin{flushright}
{\bf UCY--PHY--99/04}
\end{flushright}

\vspace*{8mm}
{\LARGE \bf \noindent \mbox{Positron spectra from internal pair conversion}}

\vspace*{2mm}
{\LARGE\bf \noindent \mbox{observed in $^{238}$U + $^{181}$Ta collisions}}

%\end{center}
 
\vspace*{5mm}
{\normalsize\bf
 \noindent 
 S.~Heinz$^{1,*}$, E.~Berdermann$^2$, F.~Heine$^1$,   
 O.~Joeres$^1$, P.~Kienle$^1$, I.~Koenig$^2$, 
 W.~Koenig$^2$, C.~Kozhuharov$^2$, U.~Leinberger$^2$, 
 M.~Rhein$^2$, A.~Schr\"oter$^2$, H.~Tsertos$^{3,\dag}$
\footnotetext{\dag Corresponding author, e-mail: "tsertos@ucy.ac.cy" \\
 Dept. of Physics, Univ. of Cyprus, PO 20537, 1678 Nicosia, Cyprus}\\
}
{\normalsize\bf\noindent 
 (The ORANGE Collaboration at GSI)}

\vspace{8mm}
{\em \noindent
$^1$ Technische Universit\"at M\"unchen,  D--85748 Garching, 
Germany\\
$^2$ Gesellschaft f\"ur Schwerionenforschung (GSI), 
  D--64291 Darmstadt, Germany\\
$^3$ University of Cyprus, CY--1678 Nicosia, Cyprus\\
}

\vspace{1cm}

\begin{center}

{\bf  (Revised version, August 23, 2000)}

\end{center}

\end{titlepage}

\vspace*{15mm}
{\noindent\bf Abstract.}
We present new results from measurements and simulations
of positron spectra, originating from
\UTa collisions at beam energies close to the Coulomb barrier.
The measurements were performed
using an improved experimental setup at the double-Orange spectrometer of GSI.
Particular emphasis is put on
the signature of positrons from Internal-Pair-Conversion (IPC) processes in
the measured \p--energy spectra, following the de-excitation of 
electro\-magnetic transitions in the moving Ta--like nucleus.
It is shown by Monte Carlo simulations that, for the chosen current 
sweeping procedure used in the present experiments,
positron emission from discrete IPC transitions can 
lead to rather narrow line structures in the measured energy spectra.
The measured positron spectra do not show evidence for line structures
within the statistical accuracy achieved, although expected from the
intensities of the observed $\gamma$--transitions 
(E$_{\gamma} \sim1250-1600$ keV) and theoretical conversion
coefficients. This is due to the reduced detection efficiency for
IPC positrons, caused by the limited spatial and
momentum acceptance of the spectrometer.
A comparison with previous results, in which lines have been observed, is 
presented and the implications are discussed. 

\vspace*{10mm}
\noindent
{\bf PACS.} 14.60.Cd {\small\sf Electrons and positrons}~-- 
            23.20.Ra {\small\sf Internal pair production}~--
            25.70.Bc {\small\sf Elastic and quasielastic scattering}~--
            25.70.De {\small\sf Coulomb excitation}~--
            25.70.Hi {\small\sf Transfer reactions}~--
            29.30.Aj {\small\sf Charged particle spectrometers: electric and
                       magnetic}

\newpage
\section{Introduction}

Previous measurements performed 
at GSI in Darmstadt by the EPOS and the Orange groups have shown the
appearance of narrow (FWHM $\simeq 30-80$ keV), and unexpected, $e^+$--lines 
at energies in the range 250--400 keV in the positron spectra, obtained from
heavy-ion collisions near the Coulomb barrier \cite{Sch83,Cle84,Koe87}.
The lines were superimposed on continuous spectral 
distributions from quasi-atomic positron emission and from
nuclear positron background, which has been determined from measured low
resolution $\gamma$-spectra and theoretical IPC conversion coefficients
\cite{Tse92,Bok90}. 
No viable explanation could be found to account for these results.
Initially, the measurements were performed to investigate the decay of the
neutral vacuum of QED predicted by theory~\cite{Grei85}, a phenomenon which 
can lead to the emission of monoenergetic positrons in {\em supercritical}
collisions~\cite{Grei85}.
But it was clear that the width of the
reported \p{}--lines was too narrow as to be attributed to this effect 
and, particularly, the occurrence of similar \p{}--lines in the so-called 
{\em subcritical} collision systems~\cite{Grei85} have excluded an 
interpretation
of the lines in the frame work of spontaneous positron emission~\cite{Koe87}.

In follow-up extended e$^+$e$^-$-coincidence experiments electrons
were measured in coincidence with positrons, and 
the measured \pn{}--sum-energy spectra revealed even narrower lines
\cite{Cow86,Koe89,Sal90,Ikoe93}.
First attempts were made to interpret these lines in the context of 
the e$^+$e$^-$--decay of a previously
unknown neutral particle with a mass around 1.8 MeV/c$^2$~\cite{Cow86}.
But soon this hypothesis was definitively ruled out by subsequent 
Bhabha-scattering experiments~\cite{Tse91}. 

Only in recent experimental~\cite{Hei98,Lein98,Bau96,APEX97b} 
and theoretical~\cite{Hof96} studies,
extended investigations of Internal Pair Conversion
(IPC) as potential source of the observed \pn{}--sum-energy lines 
in very heavy collision systems at the Coulomb barrier have been
carried out. 
To be more specific, the IPC scenario has been systematically addressed 
in our investigations by studying the collision systems \UPb and \UTa. 
It could be shown that electromagnetic transitions in
one of the colliding nuclei which de-excites by emission of IPC \pn--pairs
can lead to in principle observable narrow lines in the 
corresponding Doppler-shift
corrected \pn--sum-energy spectra and in back-to-back
\pn--coincidence spectra,  as observed with the double-Orange 
spectrometer \cite{Hei98}.
The cross sections of the observed
$\gamma$--lines are typically of the order of some mb up to several 10 mb 
leading to weak \pn--sum-energy lines with cross sections of 
0.1 $\mu$b to several $\mu$b. The experimental sensitivity for
the detection of \pn{}--lines was limited to cross sections of
this order.
Based on this new experience, we could show moreover that the
general features of
most of the previously reported weak \pn--sum-energy lines 
resemble conspicuously to the IPC process. 

In the context of the investigation of IPC processes also the energy 
distributions for positrons emitted after IPC was reconsidered.
Our focus was particularly on positron spectra from several 
$\gamma$--transitions with energies between 1250 keV and 1600 keV observed 
with high resolution $\gamma$--spectroscopy
in the Ta--like nucleus \cite{Hei98}. 
The motivation of the present
work was to find out if \p{}--emission 
from discrete nuclear transitions could give
rise to lines in the \p{}--energy spectra with characteristics similar to 
previously reported \p{}--lines. 
In these former experiments line structures
with differential cross sections of the order of 0.2 $\mu$b/sr 
up to about 1 $\mu$b/sr in Doppler-shift
uncorrected \p{}--energy spectra were reported in several collision
systems~\cite{Koe87}.
Particularly in the system \UTa, two very weak \p{}--lines at energies of 
$\sim$230 keV and of $\sim$310 keV have been observed at a beam energy of 
5.9$\times$A MeV 
with production probabilities of 
$(3.2 \pm 0.8) \times 10^{-7}/$collision [$(0.40 \pm 0.10) \mu$b/sr]
and of $(1.5 \pm 0.5) \times 10^{-7}/$collision 
[$(0.23 \pm 0.06) \mu$b/sr], respectively. 
The FWHM of the lines was (35 $\pm$ 11) keV and 
(13 $\pm$ 4) keV, respectively. Both lines appeared in coincidence with
heavy ions scattered into a rather broad angular range 
($15^{\circ}\leq \theta_{ion} \leq  50^{\circ}$)~\cite{Koe87}.

The recent systematic investigations allowed to study the response 
and sensitivity of our experimental setup
to \p{}--spectra from IPC processes. 
Additionally, extensive Monte Carlo simulations, which take into account
the progress made in the theoretical treatment of IPC in very 
heavy nuclei during the last few years~\cite{Hof96}, were performed to support 
our understanding of the IPC process. 
In these simulations we consider the complete kinematics of the collision
system \UTa at a beam energy of 6.3$\times$A MeV, as also studied in our 
experiments.  
Lepton pairs from an IPC transition are generated in the
rest frame of the emitting ion with energy distributions 
taken from theoretical calculations. 
The energies of the leptons are transformed into the
laboratory system and 
the simulated events are then analyzed with the same 
analysis program used to
analyze the experimental data. The experimental acceptance of the setup is
also considered in the simulation program.
In a second step, the laboratory energies of the leptons 
are corrected event by event by taking into account the 
angular resolution of our setup by a Monte Carlo procedure.

The experiments were performed at the UNILAC accelerator of GSI, using 
an improved experimental setup at the double-Orange spectrometer. 
The experimental setup, the methods used as well as the Doppler-shift 
technique exploited have been described in details in our previous 
publications (see e.g. in \cite{Ikoe93,Hei98,Lein98}).
In particular, the recent investigations of the collision system \UTa
were first presented and 
discussed extensively in~\cite{Hei98}, with emphasis on the appearance of
weak \pn--sum-energy lines due to IPC processes. 
Here we report the results from a complementary analysis of this collision 
system with the objective to search for corresponding line structures in
the measured \p{}--spectra.

\vspace{0.5 cm}
\section{Experimental and simulation results}

The collision system \UTa was investigated using beams of $^{238}$U with
an energy of 6.3$\times$A MeV and 1000 $\mu$g/cm$^2$ thick $^{181}$Ta targets.
The $\gamma$-ray spectrum, measured with a Ge(i) detector at $90^{\circ}$ to
the beam direction and obtained after an event-by-event Doppler-shift 
correction to the Ta--like recoiling
ions, is shown in Fig.~1a for R$_{min}$ values between 21.4 and 24 fm.  
As can be seen, several electromagnetic transitions are excited with energies 
between 1300 keV and 1600 keV.
Their total excitation cross sections have values between a few mb and 
some 10 mb. 
These $\gamma$--transitions are obviously hitherto unknown in the 
$^{181}$Ta--nucleus~\cite{Led78}, but they were also measured by the 
EPOS~\cite{Bau96}
and APEX~\cite{APEX97b} collaborations as well as by 
Ditzel et. al.~\cite{Dit96}. 
The excitation probability, P$_{\gamma}$(R$_{min}$), was determined as
a function of the distance of closest approach, R$_{min}$, by normalizing the
$\gamma$--yield in certain R$_{min}$ intervals with the
corresponding number of elastically scattered ions. 
R$_{min}$ was derived from the measured scattering angle $\theta_{ion,CM}$ 
assuming Rutherford trajectories. 

Figure 1b shows the excitation probability
P$_{\gamma}$(R$_{min}$) as a function of R$_{min}$ for the strongest
$\gamma$--transition at E$_{\gamma} =$ (1380 $\pm$2) keV, which is
representative for all $\gamma$--lines observed between 1300 keV and
1600 keV. 
The dependence of the excitation probability on R$_{min}$ is nearly
constant in the R$_{min}$ range between 20 and 25 fm. Hence it is
significantly different from the corresponding behaviour 
of the well known low-energy $\gamma$--transitions in $^{181}$Ta due to
Coulomb excitation, as demonstrated by Fig.~1c for the case
of the 718 keV E2 transition.   
This indicates a different excitation mechanism of the high-energy 
$\gamma$--lines.

All transitions shown in Fig.~1a are accompanied by \pn{}--pair-emission
after IPC with total cross sections expected between some 0.1 $\mu$b 
and several $\mu$b, assuming transitions with multipolarities $l > 0$
and IPC coefficients of the order of $10^{-4}$.
The energy partition between the positron and the 
electron of an IPC pair is 
determined by the final state interaction with the Coulomb field
of the emitting nucleus. The emission probability for an IPC 
positron with a
definite kinetic energy is described by the energy-differential pair 
conversion coefficient d$\beta$/dE$_{\p}$ which is a
function of the nuclear charge number, the energy and the
multipolarity of the transition. 
Figure 2a shows a Monte Carlo simulation of the expected emission
probabilities for an IPC positron and the partner electron as a function 
of the \p{}--energy for the strong 1380 keV transition in the Ta--like 
nucleus which is the best candidate for an observation in our
experiments. The results in Fig.~2a are given
in the rest frame of the emitting nucleus 
taking into account the latest theoretical calculations on IPC
assuming E1 multipolarity \cite{Hof96}.
Due to the final state interaction the positron of an IPC pair is 
preferentially emitted with the highest available kinetic energy of 
E$_{\p} =$ E$_{\gamma} -$2m$_e$c$^2 =$ 358 keV
(solid curve), while for the partner electron the energy is 
complementary and given by E$_{\n} =$ E$_{\gamma} -$2m$_e$c$^2 -$ E$_{\p}$
(dotted curve). 

After transformation of the \p{}--energies into the laboratory system we
obtain the distribution shown in Fig.~2b for R$_{min} = 21.4 - 24$ fm.
If we now take into account the acceptance  
of the Orange \p{}--spectrometer at the chosen current settings, 
the distribution of
Fig.~2b is reduced to the broad (FWHM $\sim$80 keV) line structure at
$\sim$280 keV shown in Fig.~2c. 
The \p{}--spectrum was scanned by ramping
the spectrometer current up and down within a preselected
current interval in the same manner as it was done in the experiment.
The current settings for the present experiment were such chosen that the
IPC positrons with energies $>\sim$280 keV were detected with maximum
efficiency, whereas those with laboratory energies 
larger than the maximum possible center-of-mass (CM) \p{}--energy of
358 keV cannot be detected. 
This is due to the fact that positrons are only detected
in the backward hemisphere with emission angles between $110^{\circ}$
and $140^{\circ}$ relative to the beam direction (see e.g. in~\cite{Hei98}). 
The shape of Fig.~2c at energies below 280 keV is affected 
by the significantly reduced efficiency of the spectrometer in this 
region and by the low--energy cut--off at 200 keV.
It should be underlined here that the present experiment was
optimized for the detection of narrow \pn{}--lines with sum energies
around 630 keV, and not for the measurement of single positron spectra.

An event-by-event Doppler shift correction to
the events of Fig.~2c leads to the CM distribution shown in
Fig.~2d. The correction was performed by taking into account the
finite angular resolution of the positron and heavy-ion 
detectors.
Particularly, the positron emission angle was set to a constant value of
$125^{\circ}$ as also used in the analysis of the real data
(for details see Ref~\cite{Hei98}). 
In this case
we obtain a line with a FWHM of $\sim$60 keV, whose maximum is now shifted to 
an energy of $\sim$320 keV relative to the uncorrected distribution.

According to these simulation results the IPC positrons from
the 1380 keV transition should give rise to a rather narrow peak-like
contribution in the measured \p{}--spectra 
with a FWHM of 60 to 80 keV and energies of 
$\sim$320 keV and $\sim$280 keV, for the CM and 
laboratory \p{}--energies, respectively. 
The CM distribution can be reconstructed
by means of an event-by-event Doppler-shift correction~\cite{Hei98}.

To simulate the total contribution of IPC positrons from the Ta--like ion to
the measured \p{}--spectra we have still to take into account the remaining
$\gamma$--transitions observed between 1300 keV and 1600 keV, for which
pure E1 multipolarity is assumed. 
Additionally we consider for the weak $\gamma$-transitions around 1500 keV
a possible admixture of an E0 contribution to an E2 transition. 
This possibility was discussed in~\cite{Hei98} where a line structure in the 
\pn{}--sum-energy spectra, obtained in \UTa{} collisions, was found after
Doppler-shift correction to the Ta--like ions. It exhibits the characteristics
of close-lying IPC lines originating from the electromagnetic transitions
between 1500 and 1550 keV and 
appears 8 times stronger than expected from the corresponding 
$\gamma$--spectrum by assuming E2 IPC coefficients of the order of $10^{-4}$.
It should be made clear that this empirical assumption does not
influence the final results at all, and has been considered here only for a
consistent treatment with our previous results~\cite{Hei98}.
 
Including these considerations we obtain from our simulations 
the  IPC \p{}--energy distributions
shown in Fig.~3. Figure 3a shows the expected \p{}--spectrum
when an event-by-event Doppler-shift
correction to the Ta--like recoils is applied, whereas Fig.~3b shows the
corresponding \p{}--spectrum in the laboratory system.
As can be seen, in the Doppler-shift corrected 
spectrum (Fig.~3a)
the 60 keV broad line-structure at $\sim$320 keV, resulting from the 
1380 keV transition, is still clearly distinguishable.
Without Doppler-shift correction both distributions are shifted and smeared
out, such that only two low intensity narrow structures 
near 300 keV and 400 keV are still visible (Fig.~3b).

The measured \p--energy spectra together with the calculated 
IPC \p{}--energy distributions from the above discussed electromagnetic 
transitions are presented in Fig.~4.
These spectra are shown in coincidence with only one ion 
(Ta--like), scattered in the R$_{min}$ range 21.4 to 24 fm, but without 
requiring a coincidence with an electron. 
The underlying R$_{min}$ range was chosen in order to 
optimize the signal-to-background ratio for the IPC positrons.
The solid line in Fig.~4a shows the measured Doppler-shift corrected
\p{}--energy distribution while the dotted line represents the expected
distribution of the IPC positrons from the observed $\gamma$--transitions 
in the Ta--like nucleus(cf. Fig.~3a) scaled up by a factor of 10.

The expected narrow structure from the 1380 keV transition 
around 320 keV cannot be observed in the measured spectrum.
The IPC production probability from this transition is 
$(2.1 \pm 0.5) \times 10^{-7}$ per elastically scattered ion for 
R$_{min} =$ 21.4--24 fm, as calculated from the 
corresponding measured $\gamma$--transition probability and an 
IPC coefficient of $10^{-4}$. Note here that
the expected \p{}--line energy and its production probability
are very close to the values, measured for an \p{}--line 
with an energy of $\sim$310 keV in the previous experiments~\cite{Koe87}.

For our latest measurements the statistical detection limit for IPC positrons
from the 1380 keV transition is $2.6 \times 10^{-7}/$collision. 
It is extracted from the 
measured continuous spectra assuming 
a superimposed IPC \p{}--line with 
60 keV FWHM and two standard deviations statistical significance.
In this case the IPC production probability is lower than the
statistical detection limit and thus consistent with the non-observation
of the expected IPC line. 
The production probability of the continuous \p{}--distribution itself 
amounts to $(4.20 \pm 0.03) \times 10^{-6}$ per collision 
for \p{}--energies between 290 and 350 keV. It originates from unresolved
$\gamma$--transitions and collision induced atomic positrons~\cite{Tse92}.  

In the Doppler-shift uncorrected spectra (Fig.~4b) the possibility to 
observe lines from IPC positrons is even worse. The expected IPC
\p{}--distribution represented
by the dotted line in Fig.~4b (also scaled up by a factor of 10) shows only
two slight structures around 300 keV and 400 keV which are too weak as to 
be detected in the measured spectra. They are dominated by the 
continuous positron distribution.

It is worth mentioning at this point that the $\gamma$-ray spectra, 
obtained after Doppler-shift correction to the Ta--like ion, show
another pronounced structure composed of three rather close lying 
$\gamma$--lines at energies of $\sim1240$ keV, $\sim1260$ keV and 
$\sim1275$ keV (not shown in Fig 1a).
The positron production probability expected from these transitions is
$\sim1.5 \times 10^{-7}/$ collision using an IPC coefficient of
$0.5 \times 10^{-4}$. 
In the present
experiment one could expect from these transitions a $\sim50$ keV broad
\p{}--line structure centered around 235 keV in the laboratory energy spectra,
very close to the previously reported \p{}--line at $\sim$230 keV\cite{Koe87}. 
But due to the very low spectrometer efficiency ($\epsilon \sim 2 \times
10^{-3}$) for low positron energies in this experiment the expected signature 
of a 235 keV line is far below the detection limit.

Figure~5a shows the $\gamma$-ray spectrum corrected for Doppler 
shifts, assuming 
an emission from the U--like ions, scattered in the R$_{min}$ range 19--29 fm. 
Only some very weak $\gamma$--lines with energies between 1300 keV and 
1500 keV can be observed in the above R$_{min}$ range. 
For instance, one can mark two lines at (1364 $\pm$ 7) keV and 
(1402 $\pm$ 7) keV 
with differential cross sections close to 1 mb/sr.  
The corresponding IPC contribution from these
two transitions to the \p{}--energy spectrum is very small,
i.e. $\sim$$6\times10^{-9}/$collision 
for an IPC coefficient of 0.5$\times10^{-4}$.
It is obvious that this weak IPC contribution 
is undetectable in the measured spectra.
The latter are shown in Figs.~5b and Fig.~5c, 
without and after a Doppler-shift correction, respectively.
The spectra were obtained in coincidence 
with only one scattered ion (U--like) for the R$_{min}$ range from 19 to 29 fm.
From these spectra a detection limit of 
$2.4 \times 10^{-7}/$collision (2$\sigma$) is derived for an IPC line. 

\vspace{0.5cm}
\section{Summary and Conclusions}

We studied positron emission after IPC processes of several high--lying
$\gamma$--transitions in Ta-- and U--like nuclei excited in \UTa collisions 
at a bombarding energy of 6.3$\times$A MeV.
Several $\gamma$--lines have been observed in the measured $\gamma$--spectra,
taken after an event-by-event Doppler-shift
correction to the Ta--like or U--like ions, with transition energies between
1300 keV and 1600 keV. These transitions can de-excite via the IPC branch 
with IPC coefficients of the order of $10^{-4}$. 

The goal of our investigations was to gain information about the shape and
intensity of the IPC positron distributions, originating from the above
discussed $\gamma$--transitions, which can be expected in the measured
Doppler--shift corrected and uncorrected positron energy spectra.
To accomplish it, extensive Monte Carlo simulations were in addition 
carried out, particularly for the 
strongest $\gamma$--transition at 1380 keV in the Ta--like nucleus. 
They revealed that, due to the angular and
momentum acceptance of the Orange \p{}--spectrometer as used in this
experiment, IPC from these transitions can lead to rather 
narrow peak-like contributions to the \p{}--energy spectra.
The line-structures are expected at \p{}--energies between 280 and 350 keV 
and have widths
of $\sim$60 keV up to $\sim$80 keV (FWHM), in the Doppler-shift corrected 
and uncorrected energy spectra, respectively. 
Only some $\gamma$--transitions in the Ta--like nuclei appear with sufficient
strength to give rise to IPC positron lines with production probabilities
close to the detection limit. While in the U--like nucleus no transition 
was found which could give rise to any detectable positron line.

The observations discussed above are of particular interest with respect
to the appearance of weak positron lines in our previous experiments. 
More specifically, in the collision system \UTa, measured 
at a somewhat lower bombarding energy of 5.9$\times$A MeV, two lines at 
$\sim$230 keV and $\sim$310 keV with production probabilities 
of $\sim3.2 \times 10^{-7}/$collision and $\sim1.5 \times 10^{-7}/$collision, 
respectively, were reported from the Doppler-shift uncorrected 
spectra~\cite{Koe87}. 
From the present findings, an IPC origin of these \p{}--lines cannot be excluded. 
Indeed, as shown above, the 
measured $\gamma$-transitions in the Ta-like nucleus would lead to IPC \p{}--lines with
production probabilities ($\sim10^{-7}/$coll.) and energies 
($\sim$230 keV and $\sim$320 keV) which are consistent with Monte 
Carlo Simulation results based on these IPC transitions.
This does not apply, however, to the very first \p{}--lines 
found in the collision system U+U with rather high production probabilities 
($\sim10^{-5}/$coll.)~\cite{Cle84} which are not consistent with
the intensities expected from the $\gamma$-transitions measured in
this experiment.   

It should be noted again that a direct comparison between the previous 
and the present experiment is problematic due to the following facts:
First in the present experiment, the  spectrometer 
current settings were different and optimized for an electron-positron 
coincidence experiment. 
As mentioned above, in this case energies below 200 keV were completely
suppressed and the detection efficiency for positrons with energies between
200 keV and 300 keV was reduced considerably in comparison to the former
experiments. Second no high-resolution $\gamma$-ray spectra were available
in the old experiments to reveal weak discrete IPC transitions, having probably
underestimated their role at that time. 

Our recent investigations of \p{}--emission with the Orange setup, equipped
with a high-resolution $\gamma$-ray detector, revealed
that discrete nuclear transitions, populated via Coulomb excitation or in
nuclear transfer reactions, can indeed appear in the outgoing moving nuclei 
with energies above 1 MeV. IPC processes from these transitions can lead
to rather weak and narrow line-like structures in the measured
positron energy spectra.
Due to the dominating continuous \p{}-spectral distributions appearing in 
heavy-ion collisions, the 
detection of these IPC \p{}--lines is rather difficult and depends 
strongly on the special features of the setup used as well as on the 
experimental sensitivity and techniques exploited. 

Although the IPC scenario studied in the present experiment provides 
evidence for a rather simple explanation for some of our previously
reported weak \p{}--lines, it is clear that a definite proof of this scenario 
as an overall explanation of all the previously reported \p{}- and \pn{}-lines
would require new dedicated experiments, which to our knowledge are not planned 
neither at GSI nor elsewhere.
               
\vspace*{0.5cm}
{\noindent\bf Acknowledgement:}  
{\em We would like to thank all the people of the UNILAC accelerator
 operating crew for their efforts in delivering stable $^{238}$U
 beams with high intensities.}
 
\vspace*{0.5cm}
{\noindent\footnotesize  
* present address: INFN Section of Turin, I--10125 Turin, Italy and
 CERN, CH--1211 Geneva~23, Switzerland    
}

\newpage
 
\centerline{\Large \bf Figure Captions}                              
\vspace{1.0cm}
\noindent
{\bf Fig. 1.} {\bf a)} Doppler-shift corrected $\gamma$--ray spectrum
observed in the collision system \UTa at a beam energy of 6.3$\times$A MeV.  
Ta--like recoiling ions for rather peripheral collisions
(R$_{min} =$ 21.4 -- 24 fm) are assumed to be the emitter. 
Several lines appear at energies below 1600 keV 
(see also Fig.8 of Ref. \cite{Hei98}).\\ 
{\bf b)} Excitation probability of the strongest
$\gamma$--line at $\sim$1380 keV as a function of R$_{min}$. These data were
first presented in Ref. \cite{Hei98}. \\
{\bf c)} Excitation probability for a well known
E2 $\gamma$--transition at 718 keV in a lower-lying rotational band in
$^{181}Ta$ (not shown in 1a) as a function of R$_{min}$. 

\vspace{1.0cm}
\noindent
{\bf Fig. 2.} Monte Carlo simulation results for energy distributions
expected for \p{} and \n{} emission after IPC in a nucleus
with a charge number of Z$=$73 and an electromagnetic transition with an energy
of E$_{\gamma} =$ 1380 keV and E1 multipolarity.  \\
{\bf a)} Original \p{} ({\em solid line}) and \n{} ({\em dotted line}) 
energy distributions expected 
in the rest frame of the emitting nucleus at the emission time. 
The sum of their kinetic energies has always a constant
value which is given by the energy of the $\gamma$--transition. \\
{\bf b)} The original \p{}--energy distribution after transformation into
 the laboratory system to account for the Doppler shift due to the
motion of the emitting ion. \\
{\bf c)} The corresponding \p{}--energy distribution in the laboratory
system by taking into account 
the chosen momentum acceptance of the spectrometer.\\ 
{\bf d)} The same events as in 2c, but after an event-by-event 
Doppler-shift correction has been applied utilizing 
the angular resolution of the setup.

\vspace{1.0cm}
\noindent
{\bf Fig. 3.}  Monte Carlo simulation of the \p{}--energy distribution 
expected 
from several IPC transitions in Ta (Z$=$73) with energies between 
1250 keV and 1600 keV, as suggested by the $\gamma$-ray spectrum shown 
in Fig.~1a. 
All the $\gamma$--transitions are assumed to be of pure E1 multipolarity, with the
exception of the small contribution of those appearing between 1500 keV and
1600 keV for which an admixture of E2 and E0 with a mixing ratio of
1:8~\cite{Hei98} is taken into account. \\
{\bf a)}The \p{}--energies are corrected for Doppler shifts, assuming
an emission from the Ta--like ions.\\  
{\bf b)} The corresponding \p{}--energy distribution 
expected in the laboratory system is shown.

\vspace{1.0cm}
\noindent
{\bf Fig. 4.} {\bf a)} Measured \p{}--energy spectrum from \UTa
 collisions at a beam energy of 6.3$\times$A MeV.
The {\em dotted-line histogram} indicates the \p{}--contribution, multiplied by
10, which can be 
expected from IPC of excited states in the Ta--like nucleus (cf. Fig. 3a). 
Both spectra are corrected for Doppler shifts, assuming an emission from the 
Ta--like recoils scattered in 
the R$_{min}$ range between 21.4 and 24 fm. \\
{\bf b)} The corresponding spectra obtained in the laboratory system are shown.

\vspace{1.0cm}
\noindent
{\bf Fig. 5.} {\bf a)} Measured $\gamma$-ray spectrum, corrected for 
Doppler shifts, assuming an emission from the U--like ions 
scattered in the R$_{min}$ range from 19 to 29 fm. \\
{\bf b)} Measured \p{}--energy spectrum in the laboratory system for the
above R$_{min}$ range.\\ 
{\bf c)} The same as in Fig. 5b, but after Doppler-shift correction to the
         U--like scattered ions.

%----------------The figures are put here --------------------------------- 

\newpage
\pagestyle{empty}

\begin{center}
 \epsfig{file=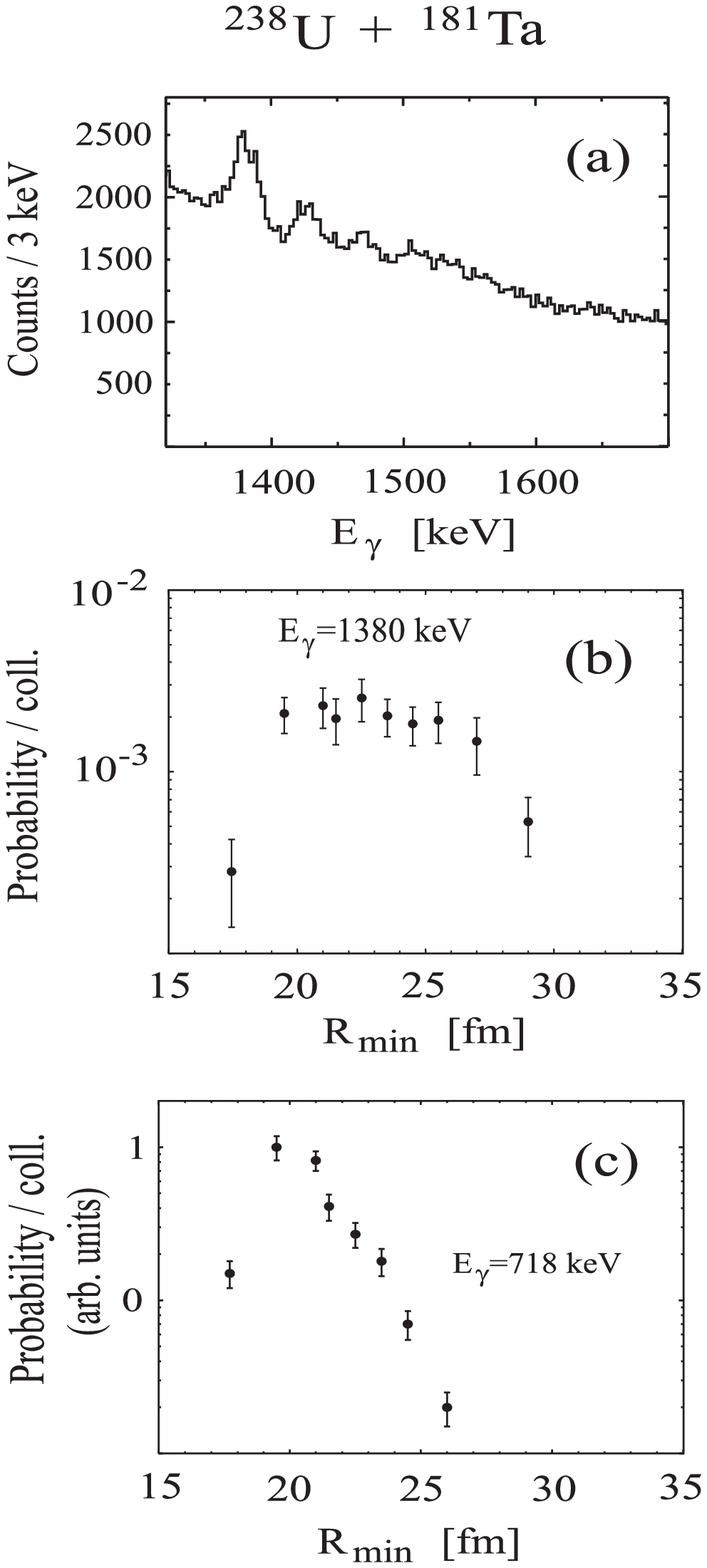,height=21cm}
\end{center}
 
\vspace*{0.3cm}
{\Large \bf Figure 1}
 
\vspace*{3 mm}
(S. Heinz {\em et al.}, EPJ A)

\newpage
\pagestyle{empty}
 
\vspace*{2.5cm}
\begin{center}
 \epsfig{file=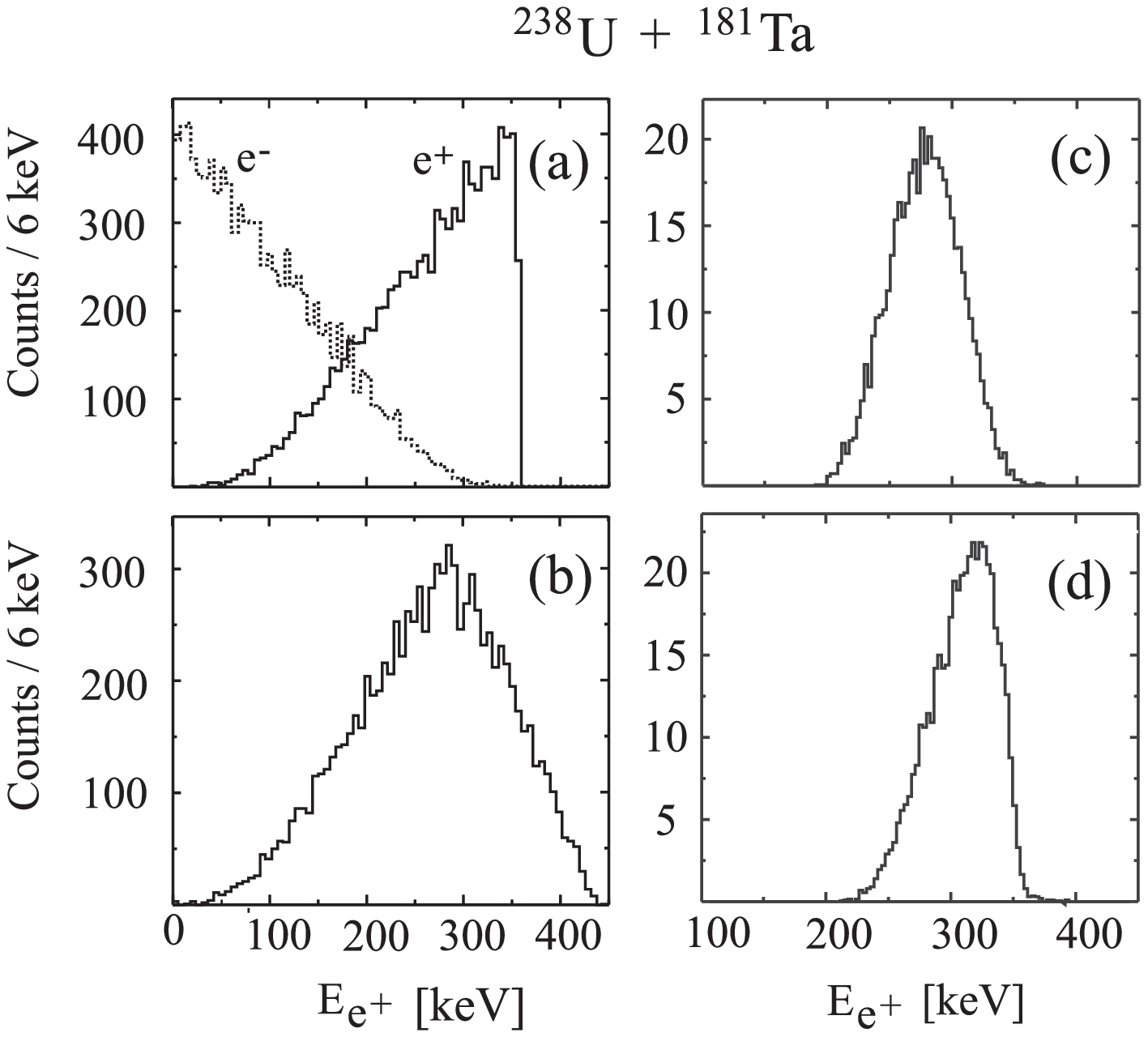,width=16cm}
\end{center}
 
\vspace*{0.3cm}
{\Large \bf Figure 2}
 
\vspace*{3 mm}
(S. Heinz {\em et al.}, EPJ A)

\newpage
\pagestyle{empty}
 
\begin{center}
 \epsfig{file=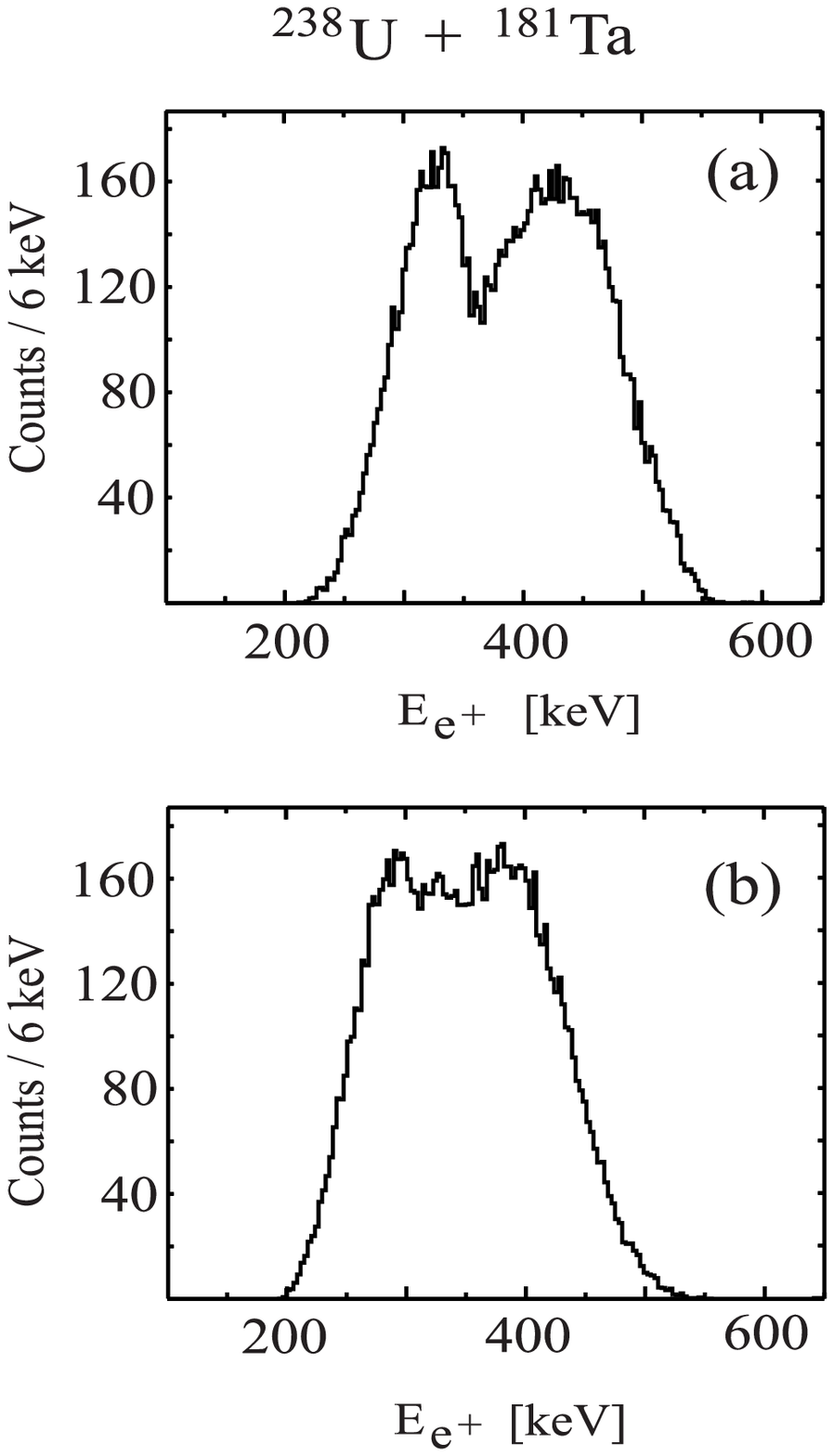,height=21cm}
\end{center}
 
\vspace*{0.3cm}
{\Large \bf Figure 3}
 
\vspace*{3 mm}
(S. Heinz {\em et al.}, EPJ A)

\newpage
\pagestyle{empty}
 
\begin{center}
 \epsfig{file=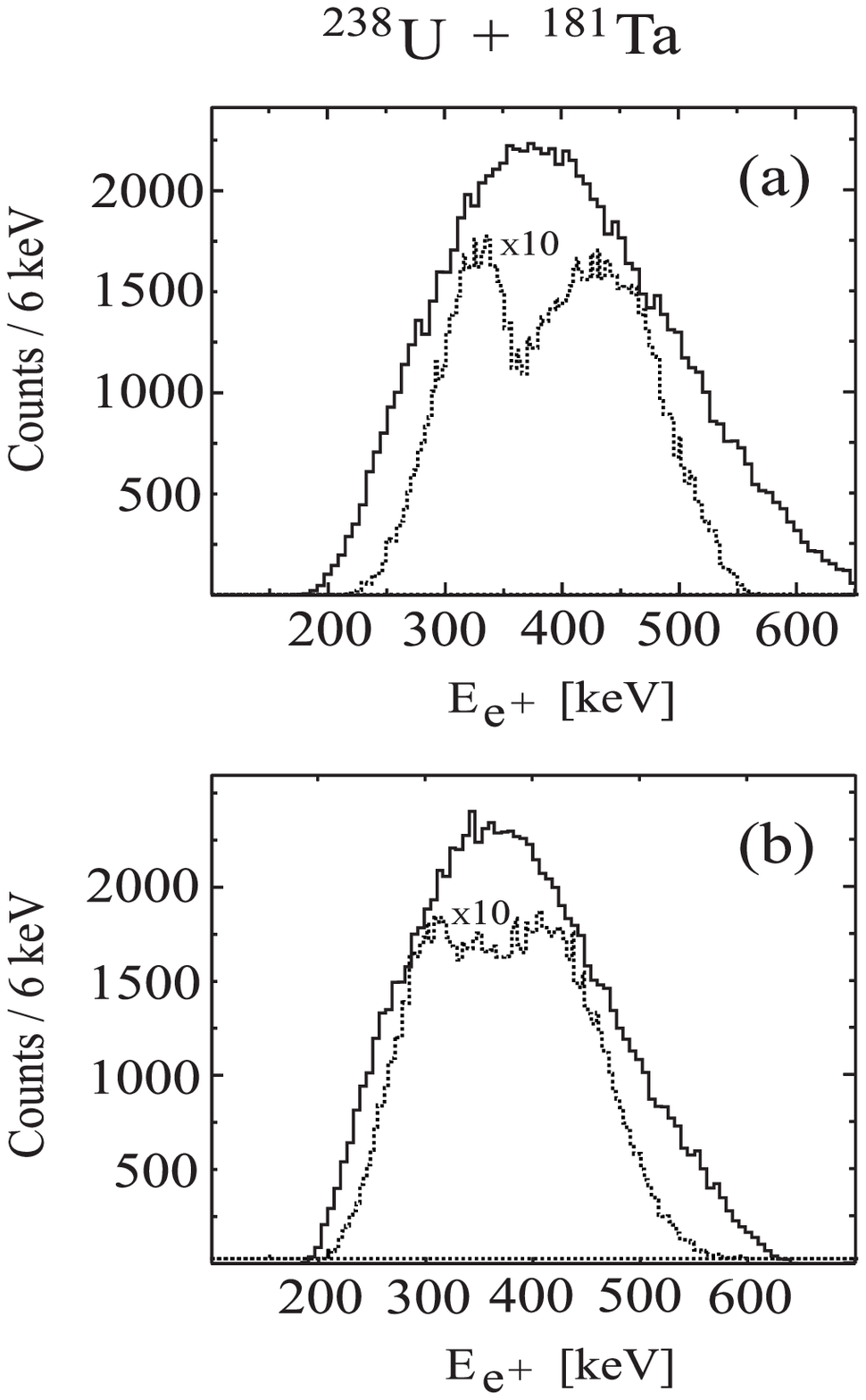,height=21cm}
\end{center}
 
\vspace*{0.3cm}
{\Large \bf Figure 4}
 
\vspace*{3 mm}
(S. Heinz {\em et al.}, EPJ A)

\newpage
\pagestyle{empty}

\begin{center}
 \epsfig{file=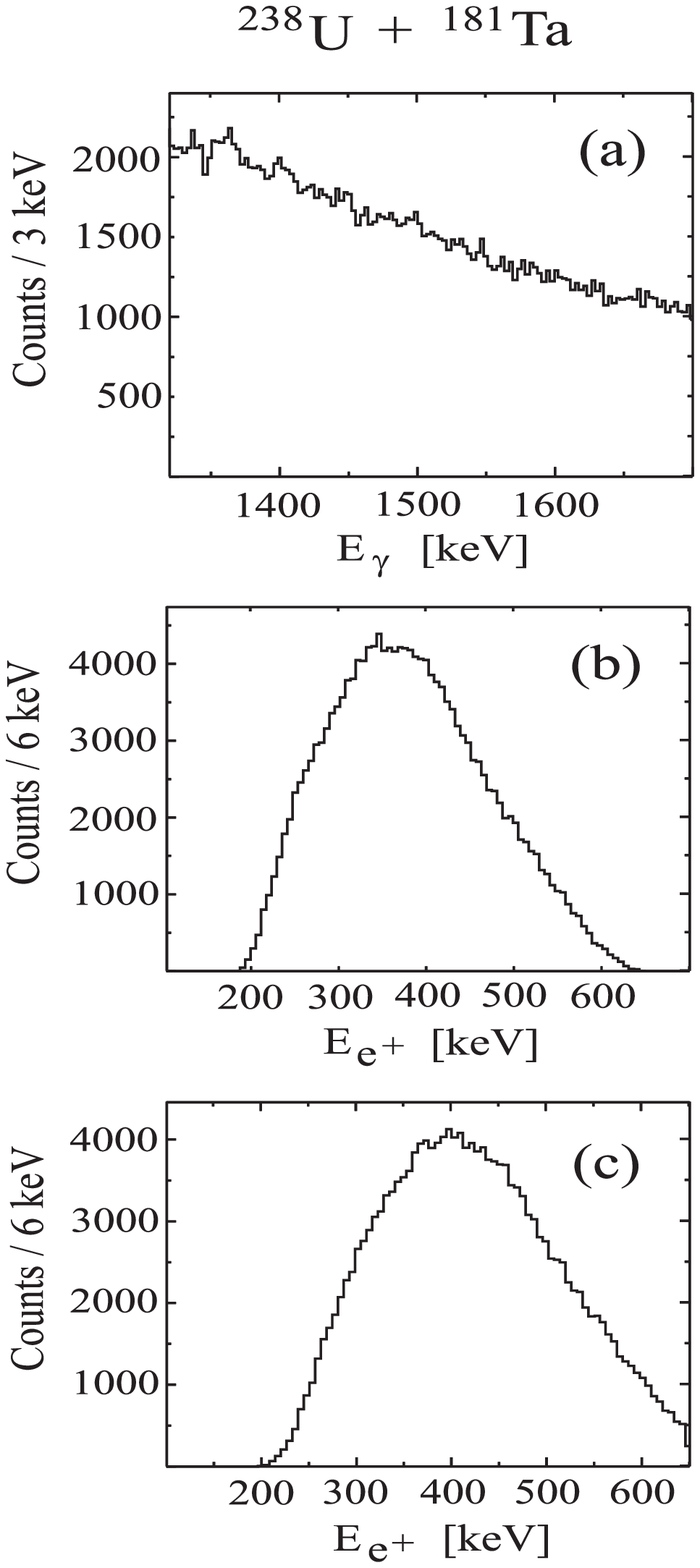,height=21cm}
\end{center}

\vspace*{0.3cm}
{\Large \bf Figure 5}

\vspace*{3 mm}
(S. Heinz {\em et al.}, EPJ A)

\end{document}